%% file: main.tex
\documentclass{article}
\usepackage[utf8]{inputenc}
\usepackage{amssymb}

\usepackage{todonotes}
\usepackage{pgfplots}
\usepackage{wrapfig}
\usepackage{amsmath}
\usepackage{indentfirst}
\pgfplotsset{width=7cm,compat=1.9}
\makeatletter
\let\@fnsymbol\@arabic
\makeatother
\graphicspath{ {./images/}}
\usepackage{float}
\usepackage[english]{babel}
\usepackage[letterpaper,top=2cm,bottom=2cm,left=3cm,right=3cm,marginparwidth=1.75cm]{geometry}
\usepackage{hyperref}
\UseRawInputEncoding

\date{}
\title{\rule[0pt]{\textwidth}{1pt}\vspace{0.5cm} Introduction to the Artificial Intelligence that can be applied to the Network Automation Journey \\ \rule[0pt]{\textwidth}{1pt}\par}
\author{Alexandre GONZALVEZ\thanks{Computer Engineering Student, INSA Toulouse.} \\ \emph{alexandre.gonzalvez@nxo.eu} \and Gilbert MOISIO\thanks{Network \& Methodology, Senior Consultant.} \\ \emph{gilbert.moisio@nxo.eu} \and Noam ZEITOUN\thanks{Project Expert, NetDevOps.} \\ \emph{noam.zeitoun@nxo.eu}}

\begin{document}

\maketitle

\begin{center}
\end{center}

\begin{abstract}
The computer network world is changing and the NetDevOps approach has brought the dynamics of applications and systems into the field of communication infrastructure. Businesses are changing and businesses are faced with difficulties related to the diversity of hardware and software that make up those infrastructures. The "Intent-Based Networking - Concepts and Definitions" document describes the different parts of the ecosystem that could be involved in NetDevOps. The recognize, generate intent, translate and refine features need a new way to implement algorithms. This is where artificial intelligence comes in.
\end{abstract}

\input{content/__all__.tex}

\bibliographystyle{IEEEtran}
\newpage
\input{bibliography.bbl}

\end{document}

%% file: content/__all__.tex
\input{content/01Introduction.tex}

\input{content/02The_Perceptron_and_activation_functions.tex}

\input{content/03Neural_Network.tex}
\input{content/04Artificial_Intelligence_in_Intent_Based_Networking.tex}

\input{content/99Conclusion.tex}

%% file: content/01Introduction.tex
\section{Introduction}
For several years, much research has been carried out in order to understand and predict the future. Those researches are generally based on statistical techniques. But there is now a new challenger to traditional statistical models: neural network models. The idea behind it is to give a computer a lot of examples of inputs and outputs, and then hope that the computer can find a way to relate the two in a meaningful way and generalizes the pattern. The way it learns a meaningful relationship is done through a series of connections between neurons. Each of these connections has a weight which represents the importance of the connection and each neuron has a bias which is a number added to the neuron to give it a higher or lower activation.
NetDevOps is in the process of having an IETF standard that describes the concept of Intent-Based Networking, in which the relationship between the User Space part and the Intent-Based System Space part is being explored. Artificial Intelligence can help to solve problems that would require very long algorithms weighed down by long test suites.
\vspace{3cm}
\par\bigskip

%% file: content/02The_Perceptron_and_activation_functions.tex
\section{The Perceptron and activation functions}
The perceptron (or a neuron) is a fundamental particle of neural networks. It works on the principle of thresholding. Let \(f(x)\) be a summation function with a threshold of 40.
\begin{figure}[H]
    \centering
    \includegraphics[width=0.5\textwidth]{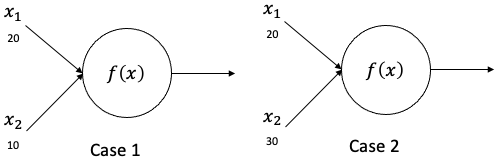}
    \caption{Firing up the neuron.}
\end{figure}
In both cases, the defined function returns the addition of two inputs, \(x_{1}\) and \(x_{2}\). In case 1, the function returns 30 which is less than the threshold value. In case 2, the function returns 50 which is above the threshold and the neuron will fire. Now this function becomes more complicated than that. A neuron in a typical neural network receives a sum of input values multiplied by its weights. Then we add the bias, and the function, also known as the activation function or step function, helps to make a decision.
\begin{figure}[H]
    \centering
    \includegraphics[width=0.8\textwidth]{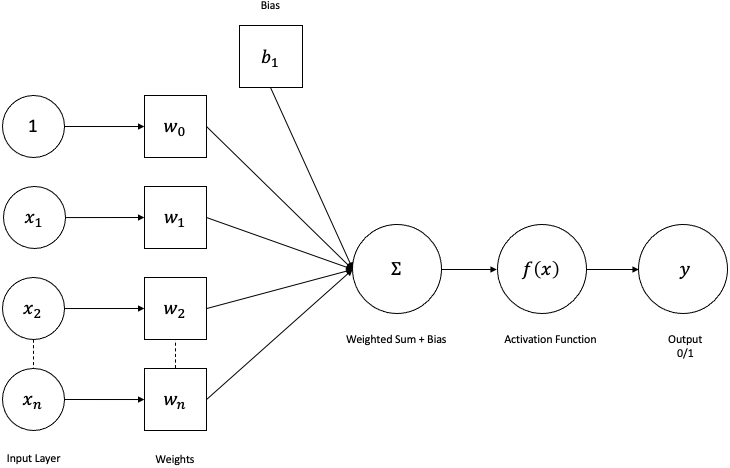}
    \caption{Perceptron.}
\end{figure}
\par\bigskip 
\noindent
\vspace{.5cm}
Activation functions convert the node's output into a binary output; 1 if the weighted input exceeds the threshold, 0 otherwise (depends on the activation function). There are three best known activation functions:
\vspace{2.5cm}
\subsection{Sigmoid and Hyperbolic Tangent (Tanh)\label{tanh}}
Sigmoid is a widely used activation function that helps capture non-linear relationships.
\vspace{.5cm}
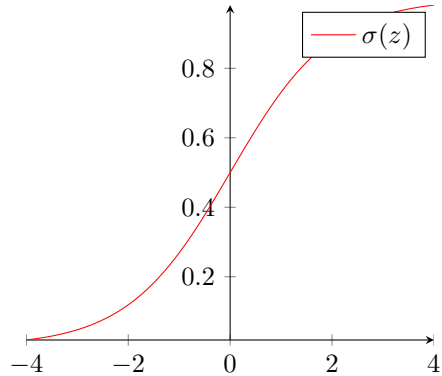
\begin{figure}[H]
\centering
\begin{tikzpicture}
\begin{axis}[
    axis x line=bottom,
    axis y line=center,
    xtick={-4,-2,...,4},
]
\addplot [
    domain=-4:4, 
    samples=100, 
    color=red,
]
{1/(1 + e^(-x))};
\addlegendentry{\(\sigma(z)\)}
\end{axis}
\end{tikzpicture}
\caption{Sigmoid curve.}
\[
   \sigma(z) = \frac{1}{1+e^{-z}}
\]
\end{figure}
\par\bigskip 
\noindent
For any value of \(z\), if the input to the function is either a very large negative number or a very large positive number, the function \(\sigma(z)\) will always return \emph{0} or \emph{1} as an output. For this reason, it is widely used in probability-based questions. 
\vspace{.5cm}
\begin{figure}[H]
\centering
\begin{tikzpicture}
\begin{axis}[
    axis x line=center,
    axis y line=center,
    xtick={-2,-1,...,2},
]
\addplot [
    domain=-2:2, 
    samples=100, 
    color=red,
]
{tanh(x)};
\addlegendentry{\(\tanh(z)\)}
\end{axis}
\end{tikzpicture}
\caption{Hyperbolic tangent curve.}
\[
   \tanh(z) = \frac{e^{z}-e^{-z}}{e^{z}+e^{-z}}
\]
\end{figure}
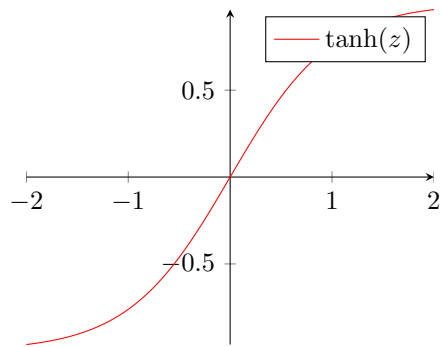
\par\bigskip 
\noindent
The Hyperbolic Tangent looks more or less like the sigmoid function but tanh varies from -1 to 1, which makes it suitable for classification problems. It is non linear.
\vspace{1cm}

\subsection{Rectified Linear Unit (ReLU) and Gaussian Error Linear Unit (GELU) \label{gelu}}
It is the most used activation function in deep learning because it is less complex than other activation functions, nonetheless efficient. \(ReLu(z)\) returns 0 or \(z\) .
\vspace{.5cm}
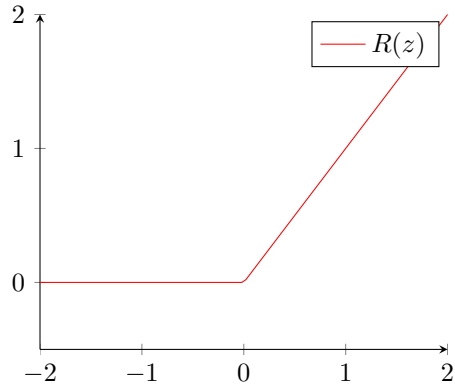
\begin{figure}[H]
\centering
\begin{tikzpicture}
\begin{axis}[
    axis x line=left,
    axis y line=left,
    xtick={-2,-1,...,2},
    ymin=-0.5,
]
\addplot [
    domain=-2:2, 
    samples=100, 
    color=red,
]
{max(0.0, x)};
\addlegendentry{\(R(z)\)}
\end{axis}
\end{tikzpicture}
\caption{ReLu curve.}
\[
    R(z)= \max{(0,z)} =
\begin{cases}
    x_i,& \text{if } x < 0\\
    0,& \text{if } x > 0
\end{cases}
\]
\end{figure}
\par\bigskip 
\noindent
This makes the calculation easier because the derivative of the function \(R(z)\) returns 0 or 1. The GELU \cite{hendrycks_gaussian_2020} is an activation function used in \emph{Google's BERT} (described in \ref{bert}) and \emph{OpenAI's GPT-2}. It is only 6 years old (2016), but receives just recently any interest.
This activation function can be written as an equation as follows:
\[
GELU(z) = z\cdot P(X \leq z) = z\cdot\phi(z) = x \cdot \frac{1}{2} \Big[ 1+erf(x/\sqrt{2})\Big]
\]
Where \(\phi(z)\) is the cumulative distribution function of the standard normal distribution.
\vspace{.5cm}
\begin{figure}[H]
\centering
\begin{tikzpicture}
\begin{axis}[
    axis x line=left,
    axis y line=left,
    xtick={-2,-1,...,2},
    ymin=-0.5,
]
\addplot [
    domain=-2:2, 
    samples=100, 
    color=blue,
]
{x*(tanh(0.112084*x^3 + 2.50663*x) + 1)};
\addlegendentry{\(GELU(z)\)}
\addplot [
    domain=-2:2, 
    samples=100, 
    color=red,
]
{max(0.0, x)};
\addlegendentry{\(R(z)\)}
\end{axis}
\end{tikzpicture}
\caption{ReLu and GELU curves.}
\end{figure}
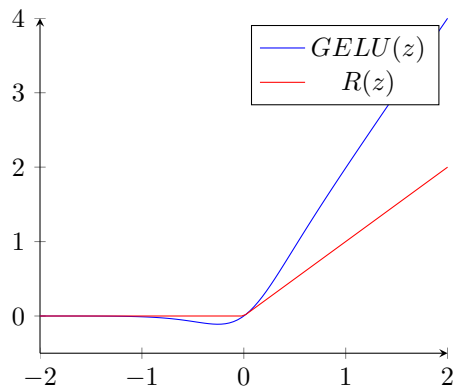
\par\bigskip 
\noindent
GELU is more interesting because it has a negative coefficient, which shifts to a \emph{positive} coefficient. So when \(x\) is greater than \(zero\), the output will be \(x\), except when \(x \in [0~;~1]\), where it slightly leans to a smaller \(y\)-value. It seems to be state-of-the-art in \emph{NLP}, specifically \emph{Transformer models}, and avoids vanishing gradients problem.

%% file: content/03Neural_Network.tex
\section{Neural Network}
To understand the black box of a neural network, let's consider a basic structure with 3 \emph{layers}; an \emph{input layer}, a \emph{hidden layer}, connected on both sides of the neurons, and an \emph{output layer}.
\begin{figure}[h]
    \centering
    \includegraphics[width=0.5\textwidth]{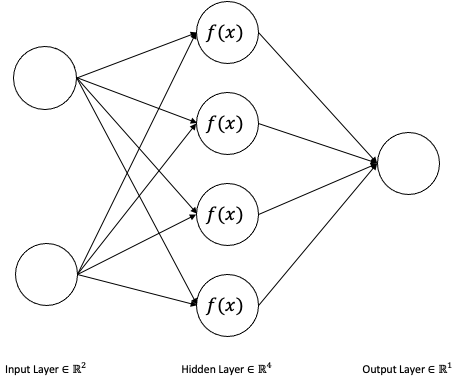}
    \caption{A simple neural network.}
\end{figure}
\par\bigskip 
\noindent
The weights and biases are randomly initialized. The accuracy of the output of a neural network consists in finding the optimal values for the weights and biases by continuously updating them. Consider an equation, \(y=wx\) where \(w\) is the weight parameter and \(x\) is the input feature. In simple terms, the weight defines the weight given to the particular input attribute (feature). Now, the solution of the equation \(y=wx\) will always pass through the origin. Therefore, an intersection is added to provide freedom to accommodate the perfect fit which is known as bias and the equation becomes \(\widehat{y}=wx+b\).\\
Therefore, the bias allows the activation functions curve to fit up or down the axis. Now let's see how complicated a neural network can become. For our network, there are two neurons in the \emph{input layer}, four in the \emph{hidden layer} and one in the \emph{output layer}. Each input value is associated with its weights and biases. The combination of input entities with weights and biases goes through the hidden layer. The network learns the entity using the activation function and it has its own weights and biases. Finally,  it makes the prediction. This is the forward propagation. The number of parameters in total for our network is \(((2\times4)+4+((4\times1)+1)=17\).

For such a simple network, a total of 17 parameters are needed to optimize to get an optimal solution. As the number of hidden layers and the number of neurons in it increases, the network gains more power, but then we have an exponential number of parameters to optimize that could end up taking up a huge amount of computing resources. So there is a trade-off.

\subsection{Cost Function}
After a single iteration of direct propagation, the \emph{error} is calculated by taking the squared difference between the actual output and the expected output. In a network, the inputs and activation functions are fixed. Therefore, we can modify the weights and biases to minimize the \emph{error}. The \emph{error} can be minimized by noticing two things: the change in \emph{error} by changing the weights by small amounts and the direction of the change.

A simple neural network predicts a value based on the linear relationship, \(\widehat{y}=wx+b\), where \(\widehat{y}\) (predicted) is the approximation to \(y\). Now, there may be several fitted linear lines \(\widehat{y}\). To choose the best-fitting line, we define the cost function.

Let \(\widehat{y}=\theta_0+x\cdot \theta_1\). We need to find values of \(\theta_0\) and \(\theta_1\) such that \(\widehat{y}\) is as close to \(y\) . To do this, we need to find values of \(\theta_0\) and \(\theta_1\) such that the following defined error is minimal.
\begin{figure}[h]
    \centering
    \includegraphics[width=0.3\textwidth]{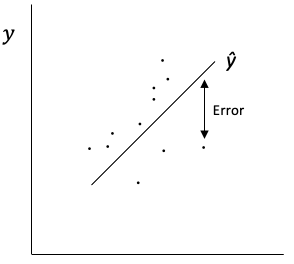}
    \caption{Best fit line.}
\end{figure}
\par\bigskip 
\noindent
The \emph{error} is the squared difference between the actual value and the predicted value which is \(E=( \widehat{y}-y)^2\). Therefore, we express the cost \(C\) with this equation 
\[
C=(1/2n)(\theta_0+x\cdot \theta_1 -y)^2
\] 
where \(n\) is the total number of points to calculate the root mean square difference and it is divided by 2 to reduce the mathematical calculation. Therefore, we need to minimize this cost function.
\subsection{Gradient Descent}
This is the algorithm that helps find the best values for \(\theta_0\) and \(\theta_1\) by minimizing the cost function. For an analytical solution and starting from \(C=(1/2n)(\theta_0+x\cdot \theta_1 -y)^2\), we take a partial differentiation of \(C\) with respect to the variables \(\theta_n\), known as \emph{Gradients}.
\begin{align*} 
\frac{\partial C}{\partial\theta} &= \frac{1}{n}\sum(\theta_0 + \theta_1 \cdot x -y), \\
\frac{\partial C}{\partial\theta} &= \frac{1}{n}\sum(\theta_0 + \theta_1 \cdot x -y)\cdot x.
\end{align*} 
These gradients represent the slope. Now the original cost function is quadratic. Thus, the graph will look like this:
\par\bigskip 
\noindent
\begin{figure}[h]
    \centering
    \includegraphics[width=0.8\textwidth]{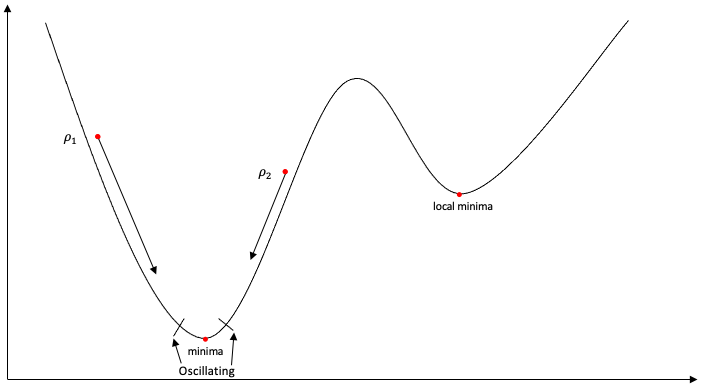}
    \caption{Gradient descent curve.}
\end{figure}
\par\bigskip 
\noindent
The equation to update \(\theta\) is:
\[
\theta^{\beta} = \theta^{\alpha} - \eta\cdot\frac{\partial C}{\partial\theta},
\]
where \(\theta^{\alpha}\) is the old one and \(\theta^{\beta}\) the new one. \\
If we are at point \(\rho1\) , the slope is negative which makes the gradient negative and the whole equation positive. Therefore, the point goes down in a positive direction until it reaches the minima. Similarly, if we are at point \(\rho2\) , the slope is positive which makes the gradient positive, and the whole equation negative moving \(\rho2\) in a negative direction until it reaches the minima. Here, \(\eta\) is the rate at which a point moves to minima known as the learning rate. All \(\theta\) are updated simultaneously and the error is calculated. Following this, we may encounter two potential problems: \\ \\
1. When updating the values of \(\theta\), you may get stuck at local minima. A possible solution is to use \emph{Stochastic Gradient Descent} (SGD) with momentum that helps to cross local minima; \\ \\
2. If \(\eta\) is too small, it will take too long to converge. Alternatively, if \(\eta\) is too large, or even moderately large, it will continue to oscillate around the minima and never converge. Therefore, we cannot use the same learning rate for all parameters. To handle this, we can program a routine that adjusts the value of \(\eta\) as the gradient moves toward the minima.

\subsection{Backpropagation}
The \emph{backpropagation} is a series of operations that optimize and update the weights and biases in a neural network using the \emph{Gradient Descent} algorithm. Consider a simple neural network (\emph{Figure 2}.) with one input, a single hidden layer and one output. \\
Let, \(x\) be an input, \(h\) be a hidden layer, \(\sigma\) be a \emph{Sigmoid} activation, \(w\) be weights, \(b\) be a bias, \(w_1\) be input weights, \(w_0\) be output weights, \(b_1\) be an input bias, \(b_0\) be an output bias, \(o\) be an output, \(E\) be an error, and \(\mu\) be the linear transformation \((\sum w_1x_1)+b\). \\
Now we create the computational graph of the \emph{Figure 2} by stacking the series of operations needed to reach from the input to the output.
\par\bigskip 
\noindent
\begin{figure}[h]
    \centering
    \includegraphics[width=0.9\textwidth]{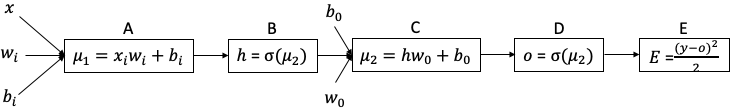}
    \caption{Computational graph.}
\end{figure}
\par\bigskip 
\noindent
Here, \(E\) depends on \(o\), \(o\) depends on \(\mu_2\), \(\mu_2\) depends on \(b_0\), \(w_0\) and \(h\), \(h\) depends on \(\mu_1\) and \(\mu_1\) depends on \(x\), \(w_1\) and \(b_1\). We need to calculate the intermediate changes with respect to weights and biases. Since there is only one hidden layer, there are input and output weights and biases. So we can divide it into two distinct cases.
\subsubsection{Case 1: Output weight and output bias.}
\par\bigskip 
\noindent
\begin{figure}[H]
    \centering
    \includegraphics[width=0.6\textwidth]{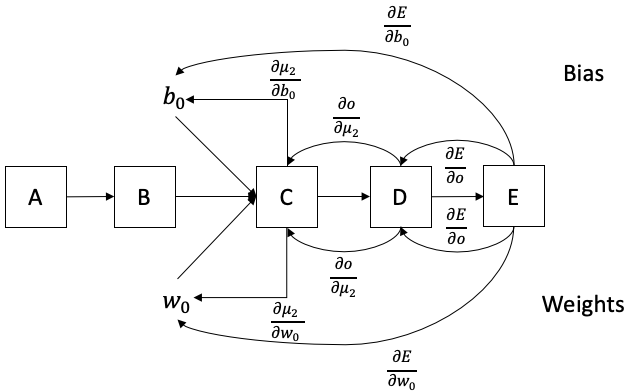}
    \caption{Computational graph for case 1.}
\end{figure}
\par\bigskip 
\noindent
By applying chain rule, following the weights path we have:
\begin{align*}
\frac{\partial E}{\partial w_0} &= \frac{\partial E}{\partial o}\frac{\partial o}{\partial \mu_2}\frac{\partial \mu_2}{\partial w_0}
\end{align*}
\text{Following the bias path we have:}
\begin{align*}
\frac{\partial E}{\partial b_0} &= \frac{\partial E}{\partial o}\frac{\partial o}{\partial \mu_2}\frac{\partial \mu_2}{\partial b_0} 
\end{align*}
\text{Therefore,}
\begin{align*}
\hspace{4.0cm} \frac{\partial E}{\partial b_0} &= \frac{2}{2}\cdot(y-o)\\&=(y-o), \\
\hspace{4.0cm} \frac{\partial o}{\partial \mu_2} &= \sigma(\mu_2)\cdot(1-\sigma(\mu_2))\\&= o\cdot(1-o), \\
\hspace{4.0cm} \frac{\partial \mu_2}{\partial w_0} &= h \text{, As \(b_0\) is constant, its derivative is 0}, \\
\hspace{4.0cm} \frac{\partial \mu_2}{\partial b_0} &= 1
\end{align*}
Where \(o(1-o)\) is the derivative of \emph{Sigmoid}. Thus, by putting the values of the derivatives in the two change equations above by mistake, we obtain gradients as follows:
\begin{align*}
\hspace{1.5cm} \frac{\partial E}{\partial w_0} &= (y-o)\cdot o\cdot (1-o)\cdot h, \\
\hspace{1.5cm} \frac{\partial E}{\partial b_0} &= (y-o)\cdot o\cdot (1-o)\cdot 1
\end{align*}
And we can update the weights and biases by the following equation:
\begin{align*}
\hspace{0.5cm} \partial w_0 &= w_0-\eta\cdot\frac{\partial E}{\partial w_0}, \\
\hspace{0.5cm} \partial b_0 &= b_0-\eta\cdot\frac{\partial E}{\partial b_0}.
\end{align*}
This calculation concerns the hidden layer and the output. Similarly, for the input and hidden layer, it is as follows with the \emph{Case 2}.
\subsubsection{Case 2: Input weight and input bias.}
\par\bigskip 
\noindent
\begin{figure}[H]
    \centering
    \includegraphics[width=0.6\textwidth]{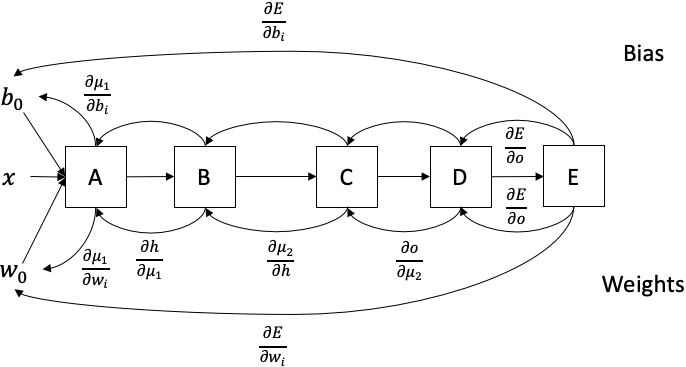}
    \caption{Computational graph for case 2.}
\end{figure}
\par\bigskip 
\noindent
\text{By applying the chain rule, following respectively the path of the weights (1) and the bias (2) we have :}
\begin{align}
\frac{\partial E}{\partial w_i} &= \frac{\partial E}{\partial o}\frac{\partial o}{\partial \mu_2}\frac{\partial \mu_2}{\partial h}\frac{\partial h}{\partial \mu_1}\frac{\partial \mu_1}{\partial w_i} \\
\frac{\partial E}{\partial b_i} &= \frac{\partial E}{\partial o}\frac{\partial o}{\partial \mu_2}\frac{\partial \mu_2}{\partial h}\frac{\partial h}{\partial \mu_1}\frac{\partial \mu_1}{\partial b_i} 
\end{align}
\text{Now, we have:}
\begin{align*}
\frac{\partial \mu_2}{\partial h} &= w_0, \hspace{2.80cm} \\
\frac{\partial h}{\partial \mu_1} &= h\cdot(1-h), \\
\frac{\partial \mu_1}{\partial w_i} &= x, \\
\frac{\partial w_i}{\partial b_i} &= 1. 
\end{align*}
\text{Therefore,} 
\begin{align*}
\frac{\partial E}{\partial w_i} &= (y-o)\cdot o\cdot(1-o)\cdot w_0\cdot h\cdot(1-h)\cdot x, \\
\frac{\partial E}{\partial b_i} &= (y-o)\cdot o\cdot(1-o)\cdot w_0\cdot h\cdot(1-h)\cdot 1, 
\end{align*}
And again, we can update these gradients using:
\begin{align*}
\partial w_i &= w_i - \eta\cdot\frac{\partial E}{\partial w_i}, \\
\partial b_i &= b_i - \eta\cdot\frac{\partial E}{\partial b_i}. 
\end{align*}
Both cases occur simultaneously and the error is calculated up to the number of repetitions called epochs. After running for a number of epochs, we have a set of optimized weights and biases for the selected features of a dataset. When new inputs to this optimized network are introduced, they are computed with the optimized values of weights and biases to obtain the maximum accuracy.

%% file: content/04Artificial_Intelligence_in_Intent_Based_Networking.tex
\section{Artifial Intelligence in the Intent-Based Networking}
This section introduces an application of artificial intelligence in Intent-Based Networking. Concretely it will present the utilization of a transformer (deep-learning model) to translate user intent in a comprehensible format for computers.
\newline
\newline
\noindent The basics of the definition of Intent-Based Networking (IBN) were published by the research group NMRG (Network Management Research Group) of the IRTF (Internet Research Task Force). This definition \cite{clemm_intent-based_2021} has evolved since 2019 and is currently at version 6, published on the 12th of December 2021. \\
The goal is to create a network accepting orders from users in the form of intent. This intent is a set of operational goals (that a network should meet) and outcomes (that a network is supposed to deliver), defined in a natural language without specifying how to achieve or implement them. \\

\begin{figure}[H]
    \centering
    \includegraphics[scale=0.45]{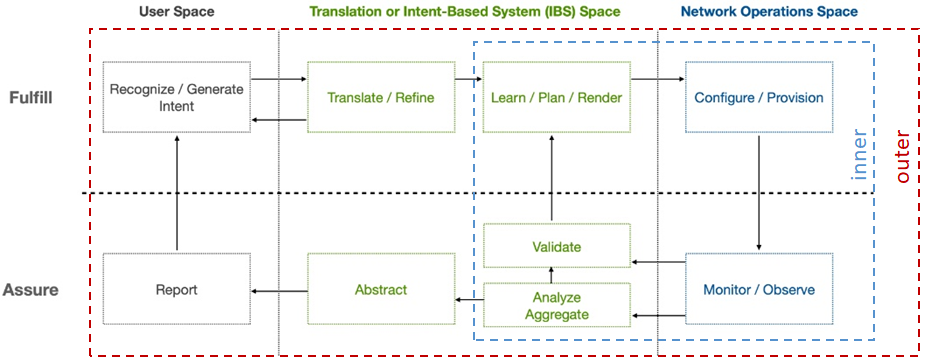}
    \caption{IBN lifecycle}
    \label{fig:ibn_lifecycle}
\end{figure}

\noindent Intent goes through a life cycle described by Figure \ref{fig:ibn_lifecycle}.

\noindent This lifecycle has 2 loops: 
\begin{itemize}
    \item The “inner” intention control loop between Intent-Based System (IBS) and Network Operations is a completely autonomous space that does not involve any human intervention. It's a Closed-loop Automation which involves automatic analysis and validation of intent-based on observations from the network operating space. 
    \item The “outer” intent control loop extends into the user space. This includes user action and adjusting their intent based on IBS observations and feedback. 
\end{itemize}

\noindent One of the most challenging tasks is to make the system understand the user intent in natural language (Recognize/Generate Intent box). Understanding natural language is a complex problem that includes the meaning of the words, sentence structure, meaning of sentences, and context. This problem involves NLP (Natural Language Processing) a subfield of computer science, linguistics, and artificial intelligence.
\newline
\newline
\noindent According to the current state, a common method to solve this problem is to divide it into two parts:
\begin{enumerate}
    \item Information extraction: Extract and label entities from the user’s input.
    \item Intent assembly: Use extracted information to recreate intent in comprehensive form for the system.
\end{enumerate}
In this paper, we will focus on information extraction. This requires part-of-speech tagging and named-entity recognition. These two NLP components can benefit from artificial intelligence progress to outperform the classical approach (ruled-based). 

\subsection{Introduction to NLP}
This section presents some helpful NLP concepts to extract information from user's requests.

\subsubsection{Structure of NLP document}

\begin{figure}[H]
    \centering
    \includegraphics[scale=0.65]{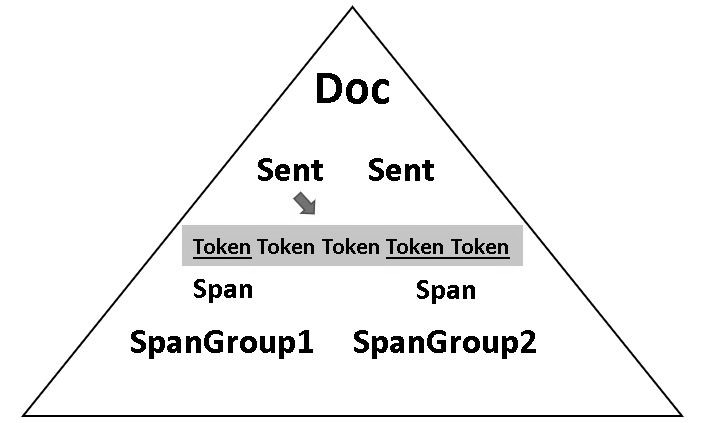}
    \caption{The classic architecture of an NLP document}
    \label{fig:struct_nlp_doc}
\end{figure}

Commonly in NLP, a document is converted into an array of sentences. Each sentence contains an array of tokens. A token is a sequence of characters that are grouped as a useful semantic unit for processing (ex: word, number, dates, acronym, punctuation). Spans are similar to tokens, in that they are a piece of a Doc container. Spans have one distinguishing feature: they can cross successive tokens. Spans can also be classified into SpanGroups. Figure \ref{fig:struct_nlp_doc} shows this hierarchy.

\subsubsection{Part-of-speech tagging and named-entity recognition}
Analyzing human speech data deeply relies on part-of-speech tagging and named-entity recognition. Those processes link tokens and spans to established categories allowing general treatment (i.e. Intent Assembly algorithms). 

\paragraph{Part-of-speech (POS) tagging}
\noindent POS tagging link each token of a text to is part of speech as shown in Table1 \ref{tab:ex_pos_tag}.

\begin{table}[H]
    \begin{tabular}{|l ||*{11}{c|}}
        \hline
        Sentence & How & many  & switches & are & up & for & more & than & 2 & hours & ? \\
        \hline
        POS Tags & SCONJ & ADJ  & NOUN & AUX & ADV & ADP & ADJ & ADP & NUM & NOUN & PUNCT \\
        \hline
    \end{tabular}
    \caption{\label{tab:ex_pos_tag}Example of POS tagging result}
\end{table}

There are mainly 2 ways of performing POS tagging:
\begin{itemize}
    \item Rules-based: Create a preset list of rules for the algorithm to follow. It's almost impossible to create enough rules to match each word of the English language to its part of speech (especially taking in consideration the position in the text).
    \item Statistical model: A statistical approach of learning to tag based on a labeled dataset. This approach can be handled by hidden Markov model, conditional random field, (deep) neural network models, or a combination of these.
\end{itemize}

\paragraph{Named-Entity Recognition (NER)} 
NER link span to a spangroup. For example instead of identifying "Barack" and "Obama' as separated entities, NER can understand that "Barack Obama" is a single entity belonging to the spangroup \textit{person}.\\
\newline
\noindent The 2 most known NER methods are the following:
\begin{itemize}
    \item Ontology-based: An ontology is a collection of data sets containing words, terms, and their interrelation. NER can rely on this knowledge base. This technique excels at recognizing known terms and concepts but the ontology needs to be extremely detailed and require updates.  
    \item Deep Learning: Recent token embedding techniques (using attention mechanisms to represent for each word their context) associated with deep learning techniques allow NER to recognize terms and concepts not present in the knowledge base.  
\end{itemize}

\subsubsection{Token embedding}
\begin{figure}[H]
    \centering
    \includegraphics[scale=0.50]{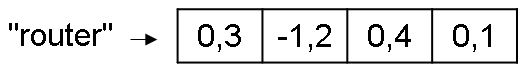}
    \caption{Example of token embedding}
    \label{fig:word_embedding}
\end{figure}

To use machine learning techniques on text data each word needs to be vectorized (as Figure \ref{fig:word_embedding}). 
\noindent Let $\phi$ be a word embedding mapping function: \\
\begin{center} $\phi$ : $\mathbb{W}$ $\rightarrow$ $\mathbb{R}^n$ \end{center} 
Where $\mathbb{W}$ is a set representing the word space and $\mathbb{R}^n$ is an n-dimensional vector space.\\

\noindent The first breakthrough was with the \textit{word2vec} model \cite{mikolov_efficient_2013} which represent each word by a specific vector taking into account word distance and word meaning (e.g. \textit{cat} and \textit{car} have close embedding considering their distance whereas \textit{spine} and \textit{switch} have a similar embedding because of their meaning). From there, other models surfaced with a vector contextualized meaning as with attention-based deep learning techniques.

\subsection{State of this art of attention-based techniques in NLP}
The context of words is essential to understand their meanin g. For example in the sentences \textit{"Switch VLAN's configuration of each device."} and \textit{"Show me the switch named spine1"} the word \textit{"switch"} have a different meaning.\\
\noindent At first bidirectional\footnote{The context of a word is given by previous and next words, that’s why a bidirectional algorithm is needed}
RNN (Recurrent Neural Network \cite{schmidt_recurrent_2019}) with attention mechanism was used to transfer all sentence information, including the relation between words\footnote{Bidirectional LSTM\cite{zhou_attention-based_2016} were deeply used and are still convenient for whole text classification}.
However, creating a tool to transform each word into a vector with meaningful information about its context does not
require RNN but only the attention mechanism.

\subsubsection{The Attention Mechanism}
\begin{figure}[H]
    \centering
    \includegraphics[scale=0.50]{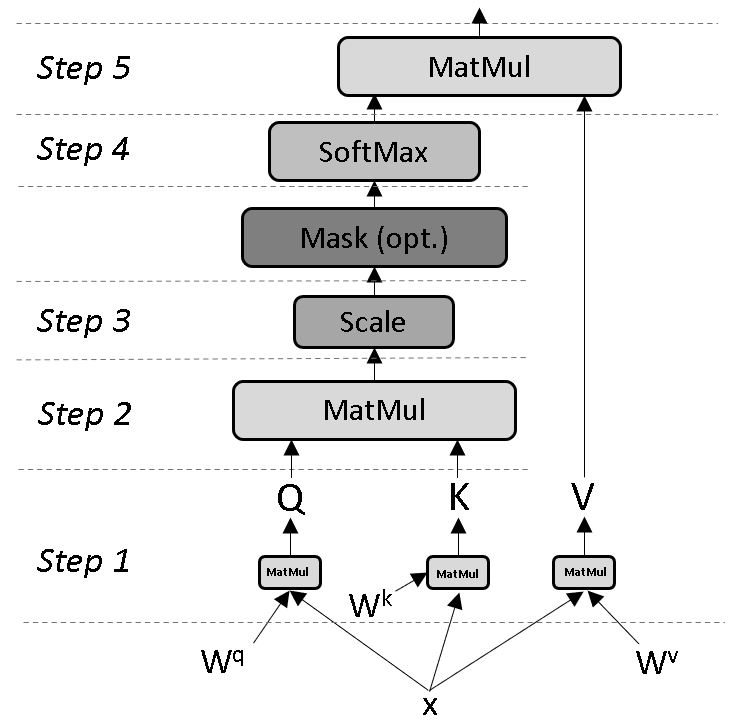}
    \caption{Scaled Dot-Product Attention}
    \label{fig:dot_product_attention}
\end{figure}

In the paper "Attention Is All You Need" \cite{vaswani_attention_2017}, they explain the Scaled Dot-Product Attention. The principle is from a matrix x (of "basic" representation of each word) to produce a new matrix A where each vector represents a word and its context. \\
\noindent This is done using 3 matrices of weights $W^q$, $W^k$, and $W^v$. These parameters are improved by backpropagation during the learning phase. \\ The structure of the scaled dot-product is shown in Figure \ref{fig:dot_product_attention}.

\begin{figure}[H]
    \centering
    \includegraphics[scale=0.40]{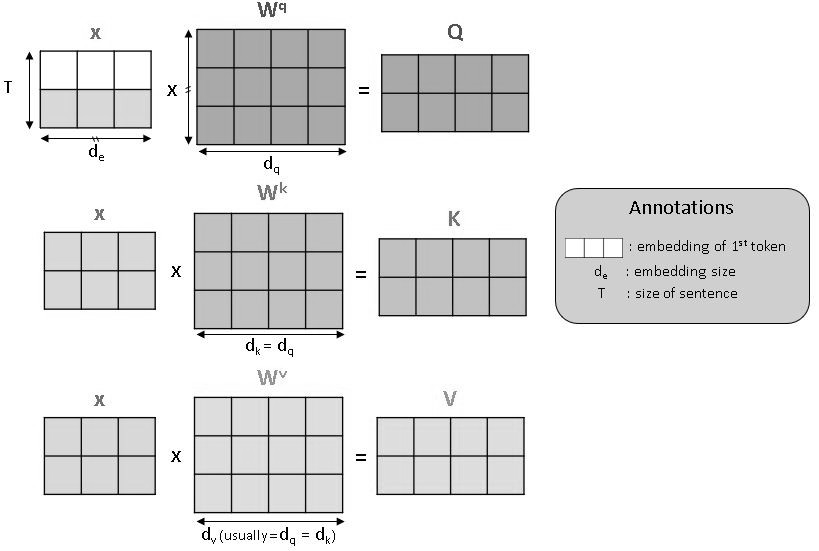}
    \caption{Calculation of Q, K and V}
    \label{fig:step1_schem}
\end{figure}

\paragraph{Step 1: Calculation of Q, K and V}
\paragraph{}
The first step is to construct queries(Q), keys(K), and values(V). This is done by matrix multiplication between x and the weighted matrices as shown in Figure \ref{fig:step1_schem}.  

\paragraph{Step 2: MatMul of $Q$ and $K^T$}
\paragraph{}
Then we construct a matrix that represent the relation between words by computing Q and $K^T$:
\begin{figure}[H]
    \centering
    \includegraphics[scale=0.40]{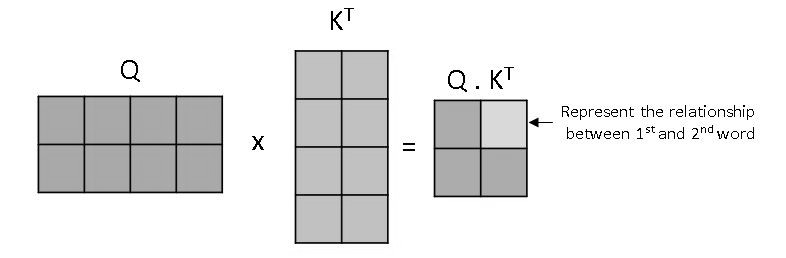}
    \caption{Matrix multiplication of Q and $K^T$}
    \label{fig:step2_schem}
\end{figure}

\paragraph{Step 3: Scaling}
\paragraph{}
To prevent softmax from being too sharp\footnote{If values given by the softmax function are to close 0 and/or 1, the gradient descent algorithm will take more steps to converge\cite{wan_influence_2019}}, we scale the product between Q and $K^T$ by dividing it by $\sqrt{d_k}$.

\paragraph{Step 4: SoftMax}
\paragraph{}
The softmax is here to normalize $\mathbb{R}$ values between 0 and 1.\\

\paragraph{Step 5: MatMul of $softmax(QK^T/\sqrt{d_k})$ and $V$}
\paragraph{}
The ending step is to multiply this relationship matrix with V to get a matrix A where each vector represents a new token embedding:
\begin{center}
$ Attention(Q,K,V) = softmax(\frac{Q.K^T}{\sqrt{d_k}}).V $
\end{center}
\begin{figure}[H]
    \centering
    \includegraphics[scale=0.40]{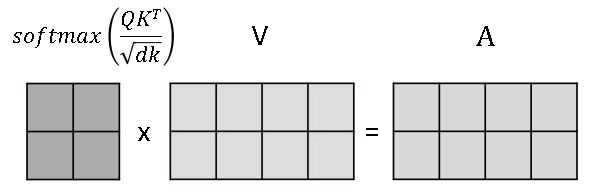}
    \caption{Matrix multiplication of $softmax(QK^T/\sqrt{d_k})$ and $V$}
    \label{fig:step5_schem}
\end{figure}
This new matrix A have T lines (1 for each token) and $d_v$ columns.

\begin{figure}[H]
    \centering
    \includegraphics[scale=0.55]{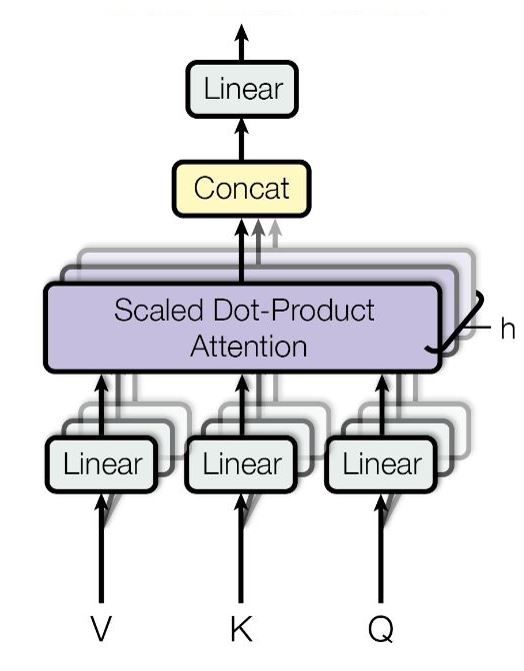}
    \caption{Multi-Head Attention - model architecture. A. Vaswani, N. Shazeer, N. Parmar, J. Uszkoreit, L. Jones, A. N. Gomez, L. Kaiser, and I. Polosukhin, Attention is all you need.}
    \label{fig:Multi-head attention}
\end{figure}

\paragraph{Multi-Head Attention}
\paragraph{}
The previous steps are representing 1-attention head. But to intent different parts in the sequence differently, we can use several\footnote{In the original paper, the model comport 8 heads} heads (similar to multiple kernels channels in CNN). Each head comports its own weighted matrices $W^q_{(i)}$, $W^k_{(i)}$, and $W^v_{(i)}$. They can be computed simultaneously. Then they are concatenated. Commonly $d_v$, the embedding size of 1 head, is equal to $\frac{d_e}{h}$ with h the number of heads, and $d_e$ the original embedding size. Therefore when concatenated the new embedding size is equal to the starting one. Finally, there is a linear layer to add more parameters to this model. The whole architecture of Multi-Head attention is represented in Figure \ref{fig:Multi-head attention}.

\subsubsection{The Transformer architecture} \label{transformer}
\begin{figure}[H]
    \centering
    \includegraphics[scale=0.55]{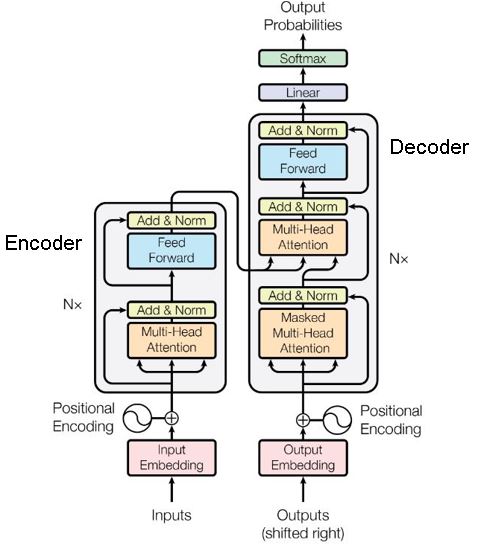}
    \caption{The Transformer - model architecture. A. Vaswani, N. Shazeer, N. Parmar, J. Uszkoreit, L. Jones, A. N. Gomez, L. Kaiser, and I. Polosukhin, Attention is all you need.}
    \label{fig:transformer}
\end{figure}

In the same paper \cite{vaswani_attention_2017}, they present transformer entire architecture (Figure \ref{fig:transformer}).
A transformer is a self-supervised model, its structure comports 2 parts: on the one hand, the encoder takes the text input and returns a representation of that input, on the other hand, the decoder part takes the expected output value (whole text) masking some of the values (including the one our model is supposed to predict, one word at a time) and returning the output probabilities of this word. The 2 last layers (Linear and Softmax) are computationally expensive because their size corresponds to the size of our dictionary(i.e. return a probability for each word of our dictionary).

\paragraph{Encoder}
\paragraph{}
To begin with, some information is added through the position of each word. As our model is not an RNN, the purpose is to make the model behave slightly differently considering the position of words in the text. It won't add exact position information, but encode relative position through 2 sinusoidal functions and add that information to the input vector.  \\
\noindent Another important behavior of this model is the skip connection part. After each layer(Multi-Head Attention layer and Feed Forward layer) there is an Add \& Norm component. Which makes the sum of the output and the input: layer(x) + x with x the layer input. This passing residual information about the data before passing through the layer. Allowing the model to let information through if the layer does not learn useful things. The normalization part is a technique to normalize the distribution of intermediate layers that enables smoother gradients, faster training, and better generalization accuracy.
\noindent Finally, the Feed Forward component is a multilayer perceptrons. It is a classic neural network where each neuron of layer n is connected to each neuron of the following layer n+1.

\begin{figure}[H]
    \centering
    \includegraphics[scale=0.55]{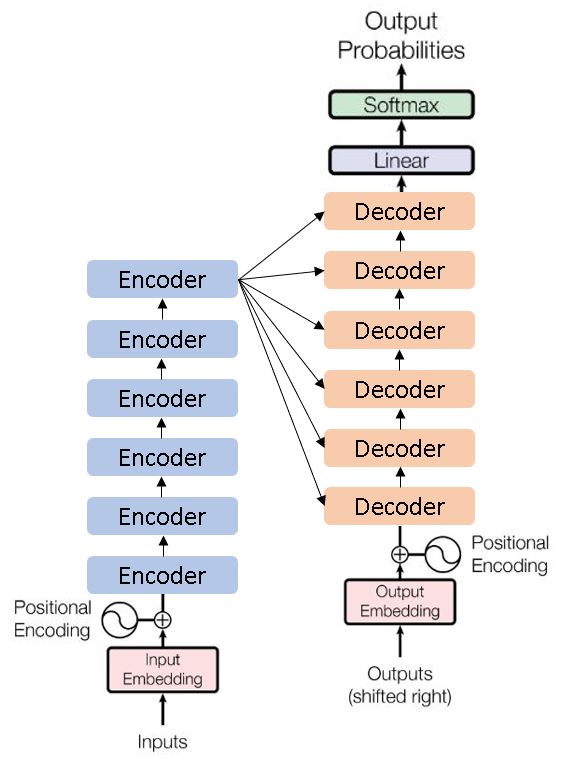}
    \caption{Computational flows with encoder and decoder repeated 6 times}
    \label{fig:n_time_transfo}
\end{figure}

\paragraph{Decoder}
\paragraph{}
On the decoder part, we have mostly the same components, the main difference is that we have Masked Multi-head attention. It is exactly the same principle, with the difference that it masks some words of the sentence as said previously. \\
\newline
\noindent Both encoder and decoder blocks are repeating N times. In the original paper, they repeat it 6 times, creating the structure represented in Figure \ref{fig:n_time_transfo}.

\subsubsection{BERT\label{bert}}
There are several implementations of the transformer model, the most popular being BERT, GPT, and BART. Their popularity is mainly due to their performance in encoding long-range dependencies through self-attention and their self-supervision techniques for leveraging unlabeled datasets. In this section, we will focus on BERT, which stands for Bidirectional Encoder Representation from Transformer \cite{devlin_bert_2019}.\\ 
\newline
\noindent There are three parts to this model: the embedding, the encoder, and the pooler layer.
\paragraph{Embedding layer\label{WordPiece_embedding}}
\paragraph{}
\noindent The BERT model's input is an array of tokens that represent a text. The token embedding is the WordPiece embeddings \cite{wu_googles_2016}. This is an example of this token embedding: 

\begin{center}
    "Show me Cisco routers up since a year" \\
     $\rightarrow$ $E_{[CLS]}$ + $E_{show}$ + $E_{me}$ + $E_{cisco}$ + $E_{routers}$ + $E_{up}$ + $E_{since}$ + $E_{a}$ + $E_{year}$ + $E_{[SEP]}$
\end{center}

\noindent This embedding allows reducing vocabulary size with generalization as \textit{verb} + \textit{ing} to "only" 30.000 different tokens. But it was created for Asian languages such as Korean and Japanese. There are a huge number of characters in these languages, as well as homonyms and no or few spaces between words. The text had to be segmented because there were no or few spaces. Without segmentation, would result in a large number of out of vocabulary terms (OOV) in the model. \\
This embedding produces a vector of size 768.

\paragraph{Encoder layer}
\paragraph{}
\noindent The encoder part is similar to the classic transformer model except for few specificity as:
\begin{itemize}
    \item The BERT model uses 12 head, so each head return vector of size 64 (768/12). 
    \item The Feed Forward layers use Gelu activation function (\ref{gelu}).  
\end{itemize}
\paragraph{Pooler layer}
\paragraph{}
\noindent The pooler uses the output representation to uses it for downstream tasks(the task you want to solve with this model). This pooler contain a linear layer and a tanh activation function (\ref{tanh}).
\newline
\paragraph{Self-supervision techniques}
\paragraph{}
\noindent This pre-trained model has learn on English Wikipedia (about 2.5 billion words)  and a book corpus (about 800 million words) with 2 tasks:
\begin{itemize}
    \item Masked Language Modeling: About 15\% of words are masked and the task is to retrieve them. It's a classification of N classes with N the size of the vocabulary (BERT vocabulary corresponds to WordPiece embedding (\ref{WordPiece_embedding}), for comparison an adult English native speaker's vocabulary is around 20 000 - 35 000 words).
    \item Next Sentence Prediction: Given 2 sentences the task is to predict if the second follows the first. It's a binary classification.
\end{itemize}
This pre-training allows the parameters of this model to be already efficient. These parameters are weights of the multi-head attention (each $Q_i,K_i,V_i$), weights, and bias of the Feed-Forward neural network.

\subsection{Using BERT to perform NER}
To personalize a pre-trained BERT model on a specific task there are 2 commons methods: both involve adding extra layers to perform our task (classification, translation, question/answer, etc.). The first called fine-tuning is to retrain the whole model, all parameters of the BERT model, and extra layers. Second, the feature-based approach is to only update the extra layers. The principle is that the last layers of the BERT model gave significant information about each word and can be used as input for our neural network.
The first method gives slightly better results but it's much more computationally expansive.

\subsubsection{Application on NER}
Applying the BERT model to specific named-entity recognition requires a few preparation steps: defining the desired span groups, creating a dataset with labeled sentences. A labeled sentence means an array containing tuples with a span group and the position of the given span (position can be indicated by the indices of the first and last token of the span or indices considering sentence character length). Then we choose to re-train the BERT model considering both methods: fine-tuning or feature-based approach.

\subsubsection{Continuous learning techniques based on user refinement}
Once trained our model can still miss some spans because they are too different by their initial embedding or by their context. In this case, by asking the user what the model miss understood, the system can correct the NER and add this new labeled sentence to the training dataset. Therefore the model will improve itself during its utilization.

%% file: content/99Conclusion.tex
\section{Conclusion}
We worked on the chapters "Intent Ingestion and Interaction with Users" and "Intent Translation" of the "Intent-Based Networking - Concepts and Definitions" document.

The first chapter of the document specifies that "The goal is ultimately to make IBSs as easy and natural to use and interact with as possible, in particular allowing human users to interact with the IBS in ways that do not involve a steep learning curve that forces the user to learn the "language" of the system. Ideally, it will be the Intent-Based Systems that is increasingly be able to learn how to understand the user as opposed to the other way round. Of course, further research will be required to make this a reality."

The second chapter indicates that "Beyond merely breaking down a higher layer of abstraction (intent) into a lower layer of abstraction (policies, device configuration), Intent Translation functions can be complemented with functions and algorithms that perform optimizations and that are able to learn and improve over time in order to result in the best outcomes, specifically in cases where multiple ways of achieving those outcomes are conceivable."

Looking at the way to answer to those two chapters, we study the way to apply Artificial Intelligence to Intent-Based Networking. We will continue to develop the concepts described is this paper in Python to finalize our Proof of Concept. All the IBN inner loop of the Intent Lifecycle, that we already developed, is currently efficient and we will work on the top part of the outer loop in order to be able to take an intent in natural language and translate it in a way understandable by the main manufacturers specialized in network automation. AI will help to create agnostic approach that a lot of people are waiting for.

%% file: bibliography.bbl

%% file: main.bbl
\begin{thebibliography}{1}
\providecommand{\url}[1]{#1}
\csname url@samestyle\endcsname
\providecommand{\newblock}{\relax}
\providecommand{\bibinfo}[2]{#2}
\providecommand{\BIBentrySTDinterwordspacing}{\spaceskip=0pt\relax}
\providecommand{\BIBentryALTinterwordstretchfactor}{4}
\providecommand{\BIBentryALTinterwordspacing}{\spaceskip=\fontdimen2\font plus
\BIBentryALTinterwordstretchfactor\fontdimen3\font minus
  \fontdimen4\font\relax}
\providecommand{\BIBforeignlanguage}[2]{{%
\expandafter\ifx\csname l@#1\endcsname\relax
\typeout{** WARNING: IEEEtran.bst: No hyphenation pattern has been}%
\typeout{** loaded for the language `#1'. Using the pattern for}%
\typeout{** the default language instead.}%
\else
\language=\csname l@#1\endcsname
\fi
#2}}
\providecommand{\BIBdecl}{\relax}
\BIBdecl

\bibitem{hendrycks_gaussian_2020}
\BIBentryALTinterwordspacing
D.~Hendrycks and K.~Gimpel, ``Gaussian error linear units ({GELUs}).''
  [Online]. Available: \url{http://arxiv.org/abs/1606.08415}
\BIBentrySTDinterwordspacing

\bibitem{clemm_intent-based_2021}
\BIBentryALTinterwordspacing
A.~Clemm, L.~Ciavaglia, L.~Z. Granville, and J.~Tantsura, ``Intent-based
  networking - concepts and definitions,'' no.
  draft-irtf-nmrg-ibn-concepts-definitions-06. [Online]. Available:
  \url{https://datatracker.ietf.org/doc/draft-irtf-nmrg-ibn-concepts-definitions}
\BIBentrySTDinterwordspacing

\bibitem{mikolov_efficient_2013}
\BIBentryALTinterwordspacing
T.~Mikolov, K.~Chen, G.~Corrado, and J.~Dean, ``Efficient estimation of word
  representations in vector space.'' [Online]. Available:
  \url{http://arxiv.org/abs/1301.3781}
\BIBentrySTDinterwordspacing

\bibitem{schmidt_recurrent_2019}
\BIBentryALTinterwordspacing
R.~M. Schmidt, ``Recurrent neural networks ({RNNs}): A gentle introduction and
  overview.'' [Online]. Available: \url{http://arxiv.org/abs/1912.05911}
\BIBentrySTDinterwordspacing

\bibitem{zhou_attention-based_2016}
\BIBentryALTinterwordspacing
P.~Zhou, W.~Shi, J.~Tian, Z.~Qi, B.~Li, H.~Hao, and B.~Xu, ``Attention-based
  bidirectional long short-term memory networks for relation classification,''
  in \emph{Proceedings of the 54th Annual Meeting of the Association for
  Computational Linguistics (Volume 2: Short Papers)}.\hskip 1em plus 0.5em
  minus 0.4em\relax Association for Computational Linguistics, pp. 207--212.
  [Online]. Available: \url{http://aclweb.org/anthology/P16-2034}
\BIBentrySTDinterwordspacing

\bibitem{vaswani_attention_2017}
\BIBentryALTinterwordspacing
A.~Vaswani, N.~Shazeer, N.~Parmar, J.~Uszkoreit, L.~Jones, A.~N. Gomez,
  L.~Kaiser, and I.~Polosukhin, ``Attention is all you need.'' [Online].
  Available: \url{http://arxiv.org/abs/1706.03762}
\BIBentrySTDinterwordspacing

\bibitem{wan_influence_2019}
\BIBentryALTinterwordspacing
X.~Wan, ``Influence of feature scaling on convergence of gradient iterative
  algorithm,'' vol. 1213, no.~3, p. 032021. [Online]. Available:
  \url{https://iopscience.iop.org/article/10.1088/1742-6596/1213/3/032021}
\BIBentrySTDinterwordspacing

\bibitem{devlin_bert_2019}
\BIBentryALTinterwordspacing
J.~Devlin, M.-W. Chang, K.~Lee, and K.~Toutanova, ``{BERT}: Pre-training of
  deep bidirectional transformers for language understanding.'' [Online].
  Available: \url{http://arxiv.org/abs/1810.04805}
\BIBentrySTDinterwordspacing

\bibitem{wu_googles_2016}
\BIBentryALTinterwordspacing
Y.~Wu, M.~Schuster, Z.~Chen, Q.~V. Le, M.~Norouzi, W.~Macherey, M.~Krikun,
  Y.~Cao, Q.~Gao, K.~Macherey, J.~Klingner, A.~Shah, M.~Johnson, X.~Liu,
  Å.~Kaiser, S.~Gouws, Y.~Kato, T.~Kudo, H.~Kazawa, K.~Stevens, G.~Kurian,
  N.~Patil, W.~Wang, C.~Young, J.~Smith, J.~Riesa, A.~Rudnick, O.~Vinyals,
  G.~Corrado, M.~Hughes, and J.~Dean, ``Google's neural machine translation
  system: Bridging the gap between human and machine translation.'' [Online].
  Available: \url{http://arxiv.org/abs/1609.08144}
\BIBentrySTDinterwordspacing

\end{thebibliography}
